\documentclass[prb,aps,preprint]{revtex4}

\usepackage{graphicx}

\begin{document}

\title{Indirect exchange coupling between magnetic adatoms in carbon
nanotubes}

\author{A. T. Costa Jr.}
\email{antc@stout.ufla.br}
\affiliation{Departamento de Ci\^encias Exatas,
Universidade Federal de Lavras, 37200-000 Lavras, MG, Brazil }
\author{D. F. Kirwan and M. S. Ferreira}\email{ferreirm@tcd.ie}
\affiliation{Physics Department, Trinity College Dublin, Dublin 2, Ireland }
\begin{abstract}

The long range character of the exchange coupling between localized
magnetic moments indirectly mediated by the conduction electrons of
metallic hosts can play a significant role in determining the magnetic
order of low-dimensional structures. Here we consider how this
indirect coupling influences the magnetic alignment of adatoms
attached to the walls of carbon nanotubes. A general expression for
the indirect coupling in terms of single-particle Green functions is
presented. Contrary to the general property that magnetic moments
embedded in a metal display Friedel-like oscillations in their
magnetic response, calculated values for the coupling across metallic
zigzag nanotubes show monotonic behaviour as a function of the adatom
separation. Rather than an intrinsic property, the monotonicity is
shown to reflect a commensurability effect in which the coupling
oscillates with periods that coincide with the lattice parameter of
the nanotube host. Such a commensurability effect does not dominate
the coupling across semiconducting zigzag or metallic armchair
nanotubes. We argue that such a long-range character in the magnetic
interaction can be used in future spintronic devices.

\end{abstract}
\maketitle

\section{Introduction}
\label{secI}

More than a decade after the discovery of carbon nanotubes, these
nanoscale cylindrical structures are still the subject of intensive
scientific research due to their intriguing physical
properties. Significant progress has been made to explain the
intrinsic properties of nanotubes but in order to expand the
applicability of those systems we need to understand how they are
affected by the interaction with other objects. Nanotubes interacting
with magnetic foreign objects are now in focus due to the possibility
of implementing the technologically promising area of spintronics in
molecular structures. In fact, the ability to produce sizeable changes
in the conductance of a nanotube due to an applied magnetic field is
one of the driving forces in the research of magnetic properties of
carbon-based structures.\cite{alphenaar} The existence of such a
spin-valve effect has potential applications in the development of
faster, smaller and more efficient nanoscopic magneto-electronic
devices.

Due to their inherent spin asymmetry, the interaction with magnetic
foreign objects is likely to cause a spin-dependent response on the
transport properties of the combined structure.
Substrates\cite{ferreira04,cespedes04}, substitutional
impurities\cite{fazzio1}, adsorbed atoms\cite{fazzio2, fazzio3} and
nanoparticles\cite{yang03}, are some of the different magnetic foreign
objects that can interact with carbon nanotubes. Among those,
transition-metal magnetic adatoms have been reported to produce
noticeable changes in the spin-dependent electronic structure of
carbon nanotubes.\cite{yang03,fazzio2}

Besides establishing how magnetic impurities affect the electronic
structure of the nanotube, it is crucial to understand the nature of
the coupling between nearby adatoms since the transport properties of
a magnetically-doped structure depend on how the impurity moments are
oriented. Dipolar and exchange interactions are the basic mechanisms
defining the alignment of the adatoms.\cite{ze+dora} The former decays
rather quickly as the moments are moved apart whereas the latter
depends on both the dimensionality and on the nature of the
interaction. Direct exchange coupling results from the overlap between
wave functions centred at the magnetic impurities but also decays
abruptly as the distance between the impurities increases. Of indirect
nature, the exchange interaction between magnetic impurities mediated
by the conduction electrons of the non-magnetic host is known to decay
more slowly and plays an important role in determining the overall
magnetic alignment of the system.

The coupling between magnetic moments mediated by the conduction
electrons of non-magnetic materials has been widely studied, both
theoretically and experimentally. The main focus of recent work has
been on metallic multilayers\cite{iec-review}, where magnetic slabs
are separated by layers of non-magnetic metals. These systems are
typically tri-dimensional, with translational symmetry preserved in
two of the three dimensions. Many of the features of the exchange
coupling in these systems are directly linked to their geometrical
properties and therefore are expected to change appreciably for
systems of lower dimensionality such as carbon nanotubes. In
particular, it has been predicted that the indirect exchange coupling
through a one-dimensional electron gas is
long-ranged.\cite{rkky-1d} If this prediction holds for the
coupling of magnetic adatoms mediated by nanotubes, this long-ranged
magnetic interaction could lead to a degree of correlation in the
electronic potential of magnetically doped nanotubes. From the
perspective of fundamental science, such a spin-dependent correlated
disorder is certain to affect the magneto-transport response of these
structures, justifying the interest of our study. On the applied side
on the other hand, our study is justified by the fact that detailed
knowledge and, perhaps, control of this magnetic coupling opens the
possibility of building quantum logic gates in solid state
environments, reminding us of the huge potential of these structures
as future components of spintronic devices.

With this motivation, in this article we investigate the indirect
magnetic coupling between two magnetic adatoms attached to the walls
of a carbon nanotube. By studying the energetic balance of the system
as a function of the relative angle between the atomic magnetizations
we can estimate the nature and magnitude of the coupling across the
nanotube. In this way, we can obtain not only how the magnetic moments
are aligned but also how it decays as the adatoms are moved further
apart. The sequence adopted in this article is the following. We start
by deriving a closed-form expression describing the energy required to
rotate the relative magnetizations of the two adatoms. Since this
expression involves matrix elements of the single-particle Green
function of the system, it is worth presenting in the subsequent
section a general way of evaluating these matrix elements for achiral
carbon nanotubes. Results and discussions are presented in section
\ref{seciv}, followed by conclusions.

\section{Indirect Magnetic Coupling}
\label{secII}

\begin{figure}
\includegraphics[width=0.9\textwidth,clip]{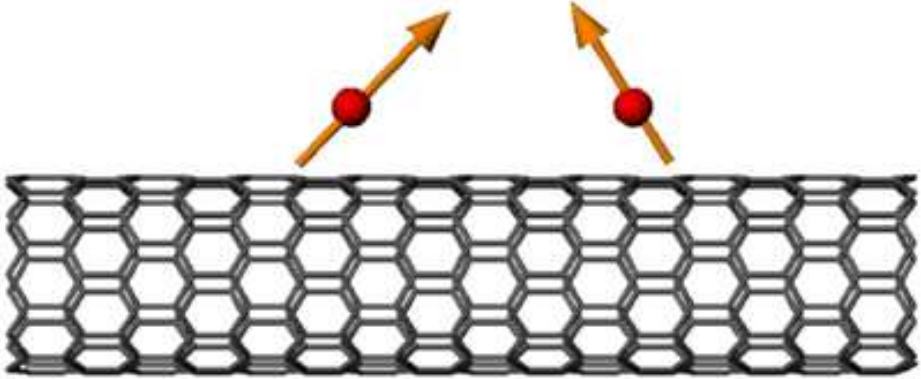}
\caption{(color online) Schematic representation of the magnetic adatoms on a carbon nanotube.}
\label{fig1}
\end{figure}

We consider two magnetic atoms, labelled $A$ and $B$, adsorbed onto
the walls of an infinitely long carbon nanotube and schematically
represented in Figure \ref{fig1}. Magnetism in these atoms is driven
by an intra-atomic Coulomb interaction that, when treated in
mean-field approximation through a self-consistent procedure, can be
described by an effective spin-dependent potential located at the
atomic positions. In this way, the electronic structure of the entire
system is well described by a single-particle Hamiltonian in a basis
of localized atomic orbitals. In such a basis, the tight-binding-like
Hamiltonian is fully determined by the on-site potentials and hopping
integrals. 

We start by assuming that the magnetic moments of the individual
adatoms are initially parallel, hereafter referred to as the
ferromagnetic (FM) configuration. In this configuration, the
Hamiltonian of the entire system written in the basis $| j \rangle$ of
atomic orbitals centred at a site $j$ is given by $\hat{H} =
\hat{H}_{NT} + \hat{H}_A + \hat{H}_B + \hat{V}_C$, where
$\hat{H}_{NT}=\sum_{j,j^\prime}|j\rangle\gamma\langle j^\prime |$ is
the Hamiltonian of the individual nanotube, $\hat{H}_A
=|a\rangle\epsilon_a\langle a |$ and $\hat{H}_B
=|b\rangle\epsilon_b\langle b |$ are the Hamiltonians associated with
the isolated atoms A and B, respectively, and $\hat{V}_C =
h\left\{|a\rangle \langle 0 | + |0\rangle \langle a | + |b\rangle
\langle n | + |n\rangle \langle b |\right\}$ refers to the coupling
between the adatoms and the nanotube. The parameters $\gamma$,
$\epsilon_a$, $\epsilon_b$ and $t$ are all matrices in spin and
orbital spaces and correspond to the hopping between nearest-neighbour
sites in the nanotube, the on-site potentials of atom A, of atom B and
the hopping between the nanotube atoms and the adatoms,
respectively. Likewise, the basis $\vert j \rangle$ represents vectors
in the same linear space. It is evident from the expressions above
that sites $j=a$ and $j=b$ label the two adatoms and $j=0$ and $j=n$
label the nanotube sites that are coupled to the magnetic atoms.

Since we are interested in evaluating the energy required to rotate
the magnetic moment of one adatom relatively to the other, it is
instructive to start by expressing the total energy of the system in
the FM configuration, {\it i.e.},
\begin{equation}
{\cal E}_{FM} = \int_{-\infty}^{\infty} d\omega \left[{\omega \over 1
+ e^{\beta (\omega-\mu)}}\right] \, \left[\left(-{1 \over
\pi}\right)\, {\rm Im} \, {\rm Tr} \sum_j
G_{j,j}(\omega)\right]\,\,\,,
\label{EFM}
\end{equation}
where $\beta=1/k_B T$, $k_B$ is the Boltzmann constant, $T$ is the
temperature, $\mu$ is the Fermi level, $G_{j,j}(\omega)$ is the Green
function of an electron with energy $\omega$ localized at site $j$ and
${\rm Tr}$ stands for the trace over the orbital and spin degrees of
freedom. To rotate the magnetic moment of one adatom, say atom $A$, we
introduce a perturbation $V(\theta)$ localized on this atom that
describes how the effective potential is altered by the rotation of an
angle $\theta$. In terms of the Pauli matrices $\sigma_x$, $\sigma_y$
and $\sigma_z$, it is given by $V(\theta) = - V_x \, \otimes \,
\left[\left(\cos\theta - 1\right)\sigma_z+\sin\theta\sigma_x\right]$,
where $V_x$ is a matrix in orbital space representing the strength of
the local exchange potentials. The energy change that results from the
perturbation $V(\theta)$ can be expressed by the respective changes
produced in the Green function $G_{j,j}$ summed over all sites $j$
(including $j=a$ and $j=b$). Making use of suitable sum rules for
Green functions, we can write this energy change as
\cite{castro94,comment}
\begin{equation}
\Delta {\cal E}(\theta) = \frac{1}{\pi}\int_{-\infty}^{\infty}d\omega
\,\left[{1 \over 1 + e^{\beta (\omega-\mu)}}\right] \, {\rm Im}\,{\rm
Tr} \, \ln [1 - G_{a,a}(\omega)\, V(\theta)]\, ,
\label{th_potential_raw}
\end{equation}
where $G_{a,a}$ represents the Green function in the FM configuration
projected at the atomic site $j=a$. Further manipulation of
Eq. (\ref{th_potential_raw}) leads to an explicit dependence of the
integrand on the spin indices, {\it i.e.},
\begin{equation}
\Delta {\cal E}(\theta) = \frac{1}{\pi}\int_{-\infty}^{\infty}d\omega
\,\left[{1 \over 1 + e^{\beta (\omega-\mu)}}\right] \, {\rm Im}\,{\rm
Tr} \, \ln [1 + 2 \, V_x^2 \, (1 - \cos\theta)
G_{a,b}^{\uparrow}(\omega)\, G_{b,a}^{\downarrow}(\omega)\,]\, ,
\label{spin-indices}
\end{equation}
where $G_{m,\ell}^{\sigma}(\omega)$ represents the propagator between
sites $j=\ell$ and $j=m$ for electrons of spin $\sigma$ and energy
$\omega$, and the trace is now only over orbital indices. Since the
$\cos\theta$ in the integrand above is the only $\theta$-dependent
term, it is reasonable to assume a Heisenberg-like angular dependence
for $\Delta {\cal E}$ and conclude that the minimum energy is either
at $\theta=0$ or $\theta=\pi$. By defining the coupling as $\Delta
{\cal E}(\theta=\pi)$ we can infer not only about the strength of the
indirect interaction between the adatoms but also whether their
moments display a parallel or anti-parallel alignment.

For the sake of simplicity the electronic structure of the system will
be here treated within the single-band tight-binding model. The
expressions above are very general and by no means restricted to such
a simple case. The results here obtained can be easily extended to a
multi-orbital description but bring no qualitative difference. Having
transition-metal atoms in mind, the adatoms are described by a 5-fold
degenerate $d$-band with the appropriate occupation to represent
typical magnetic materials. The carbon nanotube is also known to be
well described by a single-band tight-binding model with suitable
parameters that sucessfully reproduce the experimental data for the
nanotube electronic structure.  In this way, rather than matrices in
orbital indices, all quantities in the integrand of
Eq. (\ref{spin-indices}) become scalar.

It is worth mentioning that the dependence of $\Delta {\cal E}$ on the
distance between the two magnetic adatoms is contained entirely in the
matrix elements $G_{a,b}^{\sigma}$ and $G_{b,a}^{\sigma}$. To make
this dependence more explicit we can express the Green function $G$
associated with the system in the FM configuration in terms of another
Green function associated with the separate parts (also in the FM
configuration). In other words, by considering the coupling
$h$ between the nanotube and the adatoms as a perturbation, we
can use Dyson's equation to write the matrix elements of $G$ in terms
of those associated with their isolated components. It is then
straightforward to see that
\begin{equation}
G_{b,a}^\sigma(\omega) = C \,\,({ \, {\cal G}_{n,0} \over 1 - P - Q \,
{\cal G}_{n,0} \times {\cal G}_{0,n}})\,\,\,,
\label{dyson}
\end{equation}
where $C = {{\cal G}_{a,a} \, h^2 \, {\cal G}_{b,b} \over 1 - {\cal
G}_{a,a} \, h^2 \, {\cal G}_{0,0}}$, $P = {\cal G}_{b,b} \, h^2 \,
{\cal G}_{n,n}$ and $Q = {{\cal G}_{b,b} \, h^3 \over 1 - {\cal
G}_{a,a} \, h^2 \, {\cal G}_{0,0}}$. Since the Green function ${\cal
G}$ represents the electronic propagator associated with the nanotube
and adatoms in isolation, it is evident that ${\cal G}_{a,a} = {1
\over \omega - \epsilon_a}$ and ${\cal G}_{b,b} = {1 \over \omega -
\epsilon_b}$. Regarding the Green functions ${\cal G}_{0,n}$ and
${\cal G}_{n,0}$, these are the only matrix elements associated with
impurity-free nanotubes that carry the spatial dependence of the
indirect magnetic coupling between the adatoms. The indirect character
of the coupling is evident in the expressions above when written in
terms of the electronic propagators between the sites that are coupled
to the magnetic adatoms. Put in another way, the presence of those
off-diagonal propagators in the expression for the coupling indicates
that the magnetic information between the moments is being carried
back and forth by the conduction electrons of the nanotube host.
Since these propagators play such an important role in determining the
most energetically favourable magnetic configuration, in the following
section we present a fully analytical method providing closed-form
expressions for arbitrary matrix elements of the nanotube Green
function, among which the elements ${\cal G}_{0,n}$ and ${\cal
G}_{n,0}$.

\section{Green function of carbon nanotubes}
\label{secIII}

Carbon nanotubes are fullerene-like structures that can be regarded as
graphene sheets wrapped up in cylindrical shape. The electronic
structures of both graphene and carbon nanotubes are well described by
a tight-binding model for the $\pi$-orbital, the only difference
between them being the wave vector quantization along the
circumferential direction of the tubes. With such a simple dispersion
relation describing the band structure of pure tubes, one can obtain
the corresponding single-particle Green functions without having to
evaluate them numerically. In this section we derive an analytical
closed-form expression for a general matrix element of the
single-particle Green function of an infinitely long carbon
nanotube. The propagator between any two sites of an impurity-free
nanotube is obtained and will be used to calculate the indirect
exchange coupling between two magnetic adatoms.

The tight-binding Hamiltonian for a carbon nanotube is diagonal when
written in the basis $\vert {\bf k} , \pm \rangle$ defined as
\begin{equation}
\vert {\bf k} , \pm \rangle = \sqrt{{1 \over 2 L}}\sum_\ell\left[e^{i
{\vec k}.{\vec R}_\ell^\bullet} \vert \ell,\bullet\rangle \pm
e^{-i\phi({\vec k})} e^{i {\vec k}.{\vec R}_\ell^\circ} \vert
\ell,\circ\rangle \right]\,\,.
\label{muda-base}
\end{equation}
The basis $\vert \ell,\star \rangle$ represents atomic orbitals
centred at a site located at ${\vec R}_\ell^\star$ that is labelled by
a pair of indices $(\ell,\star)$, where $\ell$ is the two-atom cell
index and $\star = \bullet$ or $\star = \circ$ represent the two
inequivalent atomic sites of the hexagonal lattice. The quantity $L$
represents the total number of cells and the phase $\phi({\vec k})$ in
Eq.(\ref{muda-base}) is defined as $\phi({\vec k}) = \rm{Im} \ln {\cal
H}$, where $\rm {\cal h} = \gamma [ e^{i k_x a \sqrt{3}/3} + 2
\cos(k_y a/2) e^{-i k_x a \sqrt{3}/3} ]$. The corresponding
eigenvalues $E_{\pm}$ are found due to the basis formed by two
inequivalent carbon sites in the graphene structure. The band
structure becomes 1-dimensional when quantization conditions on the
wave vectors $k_x$ and/or $k_y$ are introduced. For achiral tubes,
namely armchair and zigzag, the wave vectors $k_x$ and $k_y$ are
quantized, respectively. In this case, the non-quantized wave vector
runs continuously along the first Brillouin zone, whereas the
quantized counterpart is only allowed to assume discrete values. For
chiral tubes in general, a linear combination of $k_x$ and $k_y$
imposes the quantization condition.  The resulting one-dimensional
energy dispersions for the armchair and zigzag nanotubes are given
by\cite{cn-book}
\begin{equation}
E_{\pm }^{j}(k_y)=\pm \gamma \sqrt{1+4\cos
(\frac{j \pi}{M} )\cos (k_y\frac{a}{2})+4\cos
{}^{2}(k_y\frac{a}{2})}
\label{armchair}
\end{equation}
and
\begin{equation}
E_{\pm }^{j}(k_x)=\pm \gamma \sqrt{1+4\cos(\frac{\sqrt{3}k_x\,a} {2}
)\cos (\frac{j \pi}{M})+4\cos^2(\frac{j \pi}{M})}\,\,,
\label{zigzag}
\end{equation}
respectively, where $M$ depends on the geometry of the tube and is
directly related to its diameter. The integer $j=0,1,2,...,M$ labels
the different bands. The nearest-neighbour electronic hopping is
denoted by $\gamma$ and is hereafter considered to be our energy
unit. Notice that the bands are symetrically distributed around $E=0$.

In what follows we calculate an arbitrary matrix element of the Green
function associated with both armchair and zigzag nanotubes, whose
band structures are given by Eqs.(\ref{armchair}) and
(\ref{zigzag}). Green functions for chiral tubes are also possible to
be obtained within the same formalism, but the corresponding unit
cells can be very large. The general formula for the single-electron
Green function is
\begin{equation}
{\cal G}(\omega )=\sum_{\vec k} \left[ \frac{\vert {\vec k},+ \rangle
\langle {\vec k},+ \vert}{\omega -E_{+}({\vec k})}+ \frac{ \vert {\vec
k},- \rangle \langle {\vec k},- \vert}{\omega -E_{-}({\vec
k})}\right] \,,
\label{General}
\end{equation}
where $\omega$ is the energy. Introducing Eq.(\ref{muda-base}) into
Eq.(\ref{General}), we obtain an expression for the Green function
between any two sites $\vert j,\star \rangle$ and $\vert
j^\prime,\star^\prime \rangle$ of a graphene sheet.  For simplicity,
we consider the propagation between two equivalent sites in different
cells. In this case the matrix element of the Green function is given
by
\begin{equation}
\langle \ell,\star \vert {\cal G}(\omega) \vert \ell^\prime,\star
\rangle = {1 \over 2 L} \sum_{\vec k} \left[ {e^{i {\vec k}.({\vec
R}_\ell^\star - {\vec R}_{\ell^\prime}^\star)} \over \omega -
E_+({\vec k})} + {e^{i {\vec k}.({\vec R}_\ell^\star - {\vec
R}_{\ell^\prime}^\star)} \over \omega - E_-({\vec k})}\right]\,\,.
\label{black-black}
\end{equation}

\subsection{Armchair nanotubes}

The sum over ${\vec k}$ in Eq.(\ref{black-black}) can be transformed
into an integral over $k_y$. The wave vector $k_x$ remains discrete
and its values are $k_x = {j \over M} {\pi \over \sqrt{3}a}$. In
addition, by taking into account that $E_-({\vec k}) = - E_+({\vec
k})$, the expression above is rewritten as
\begin{equation}
{\cal G}_{\ell,\ell^\prime}(\omega )=\frac{a}{4 M \pi }\sum_{k_x}
\int_{-\pi /a}^{\pi /a}dk_{y}[\frac{\omega \,\,e^{i {\vec k}.({\vec
R}_\ell^\bullet - {\vec R}_{\ell^\prime}^\bullet)} }{\omega
^{2}-E_{+}^{2}(k_{x},k_{y})}]\,,
\label{above}
\end{equation}
For simplicity, in Eq.(\ref{above})
we have dropped the bra and ket representing the initial and ending
sites. For each value of $k_{x}$ we have a corresponding
one-dimensional band structure. The integral above can be evaluated by
extending $k_{y}$ to the complex plane and changing the integration
contour from a straight line on the real axis to the boundaries of a
semi-infinite rectangle in the upper half-plane whose base lies on the
real axis between $-\pi /a$ and $\pi /a$. By
determining the poles and their respective residues we can evaluate
the integral for each band index $k_{x}$.  The poles, labelled
$q_j^{\pm }$, are solutions of the following equation
\begin{equation}
\cos (q_j^{\pm }\frac{a}{2})=-\frac{1}{2}\{\cos (\frac{j}{M}\frac{\pi
}{2})\pm \sqrt{\frac{\omega ^{2}}{t^{2}}-\sin ^{2}(\frac{j}{M}
\frac{\pi }{2})}\}\,,
\label{poledef}
\end{equation}
and the corresponding residues are
\begin{equation}
{\rm Res}[q_j^{\pm}]=\frac{\omega \,\, e^{i {\vec K_j}^\pm.({\vec
R}_\ell^\bullet - {\vec R}_{\ell^\prime}^\bullet)} }{4\pi M
t^{2}\{\cos (\frac{j}{M}\frac{ \pi }{2})\sin (q_j^{\pm
}\frac{a}{2})+\cos (q_j^{\pm }\frac{a}{2})\sin
(q_j^{\pm}\frac{a}{2})\}},
\label{res}
\end{equation}
where ${\vec K_j}^\pm = ({j \pi \over \sqrt{3} M a}) {\vec e}_x +
(q_j^{\pm }) {\vec e}_y$. Eq.(\ref{poledef}) does not define the poles
$q_j^{\pm}$ uniquely. By defining the cosine of pole $q_j^{\pm}$ but
not its sine, the residue in Eq.(\ref{res}) is free to assume two
distinct values. The residue depends on the correct sign of $\sin
(q_j^{\pm}\frac{a}{2})$, which is obtained by imposing that the
imaginary part of diagonal elements of retarded Green functions must
be negative. The expression for ${\cal G}_{\ell,\ell^\prime}(\omega )$
then becomes
\begin{equation}
{\cal G}_{\ell,\ell^\prime}(\omega )=\sum_j \frac{i\omega \,\, e^{i
{\vec K_j}^\pm.({\vec R}_\ell^\star - {\vec R}_{\ell^\prime}^\star)}
}{M t ^{2} \{\cos (\frac{j}{M} \frac{\pi }{2})\sin
(q_j^{\pm}\frac{a}{2})+\cos (q_j^{\pm}\frac{a}{2})\sin
(q_j^{\pm}\frac{a}{2})\}}.
\label{gwexp}
\end{equation}
This expression gives the matrix elements for the Green function of an
armchair nanotube. Although the derivation of Eq.(\ref{gwexp}) assumed
equivalent sites in distinct cells, the expression is almost the same
for inequivalent sites, except for a multiplying phase factor
$e^{-i\phi({\vec K}^\pm)}$.

It is interesting to look at the physical significance of those
poles. Each value of $k_x$ yields a corresponding $k_y$ and since they
are obtained from the band structure for a given energy $\omega$, they
are just the coordinates of constant-energy surface along the axial
direction. In other words, they indicate the wave vector $q_j^\pm$ with
which electrons of energy $\omega$ propagate along the axial
direction of the nanotube. Many discrete values of $k_x$ have no real
$k_y$ components, and therefore $q_j^\pm$ assumes imaginary values
leading to evanescent contributions to the Green functions.

\subsection{Zigzag nanotubes}

In the case of zigzag tubes the $k_y$-components are quantized whereas
the $k_x$ run continuously over the Brillouin zone. The expression for
the Green function becomes
\begin{equation}
{\cal G}_{\ell,\ell^\prime}(\omega )=\frac{a \, \sqrt{3}}{2 M \pi
}\sum_{k_y} \int_{-\pi /a\sqrt{3}}^{\pi /a\sqrt{3}}dk_{x}[\frac{\omega
\,\,e^{i {\vec k}.({\vec R}_\ell^\bullet - {\vec
R}_{\ell^\prime}^\bullet)} }{\omega ^{2}-E_{+}^{2}(k_{x},k_{y})}]\,,
\label{zig-gf}
\end{equation}

Analogously to the armchair case, the integral above can be evaluated
by indentifying the poles and summing over the respective
residues. The poles, labelled $q_j$, are obtained through the
following equation 
\begin{equation} \cos (q_j \frac{a \sqrt{3}}{2})=\frac{{\omega^2 \over
\gamma^2}-1 - 4 \cos^2(\frac{\pi j}{M})}{4 \cos(\frac{\pi j}{M})}
\label{zig-poledef} 
\end{equation} 
and the corresponding residues are
\begin{equation} 
{\rm Res}[q_j]=\frac{\omega \,\, e^{i {\vec K_j}.({\vec
R}_\ell^\bullet - {\vec R}_{\ell^\prime}^\bullet)} }{4\pi M t^{2}\cos
(\frac{\pi j}{M})\sin (q_j\frac{a \sqrt{3}}{2})},
\label{zig-res} 
\end{equation} 
where ${\vec K_j} = q_j \, {\vec e}_x + ({2 \pi j \over M a}) {\vec
e}_y$. Finally, the expression for ${\cal G}_{\ell,\ell^\prime}(\omega
)$ is given by
\begin{equation} 
{\cal G}_{\ell,\ell^\prime}(\omega )=\sum_j \frac{i\omega \,\, e^{i
{\vec K_j}^\pm.({\vec R}_\ell^\bullet - {\vec
R}_{\ell^\prime}^\bullet)} }{M t ^{2} \{\cos (\frac{j}{M} \frac{\pi
}{2})\sin (q_j^{\pm}\frac{a}{2})+\cos (q_j^{\pm}\frac{a}{2})\sin
(q_j^{\pm}\frac{a}{2})\}}.
\label{zig-gwexp} 
\end{equation} 
This expression gives the matrix elements for the Green function of a
zigzag nanotube. 

Despite having presented closed-form analytical expressions for the
Green functions of armchair and zigzag nanotubes separately, common
features in the expressions must be highlighted. The energy dependence
is not only given by the energy on the numerator of Eqs.(\ref{gwexp})
and (\ref{zig-gwexp}), but it is implicitly contained in the poles, as
shown in Eqs.(\ref{poledef}) and (\ref{zig-poledef}). Likewise, the
$M$-dependence is also implicit in the poles. Each term of the
summation corresponds to a band contribution to the single particle
Green function. One major advantage of an expression for the Green
function that does not depend on numerical evaluation, is that it
allows a transparent analysis of the relationship between the
electronic structure and the relevant parameters involved. For
diagonal elements of the Green functions, $j=j^\prime$ vanishes the
argument of the exponential in the numerator of Eqs.(\ref{gwexp}) and
(\ref{zig-gwexp}) and the imaginary part gives a direct expression for
the density of states for both armchair and zigzag nanotubes.

\section{Results and discussion}
\label{seciv}

The indirect magnetic coupling can now be calculated by inserting the
required matrix elements of ${\cal G}$ into Eq.(\ref{dyson})
and evaluating the corresponding energy integral of
Eq.(\ref{spin-indices}). By changing the relative positions between
the adatoms we can see how the coupling depends on their
separation. Furhermore, by changing the nature of the adatoms we can
also investigate how the coupling responds at the presence of
different magnetic entities. We can also investigate how the coupling
behaves for different types and sizes of nanotubes and finally, how it
changes with temperature.

\subsection{Zigzag (metallic)}

\begin{figure}
\includegraphics[width=0.9\textwidth,clip]{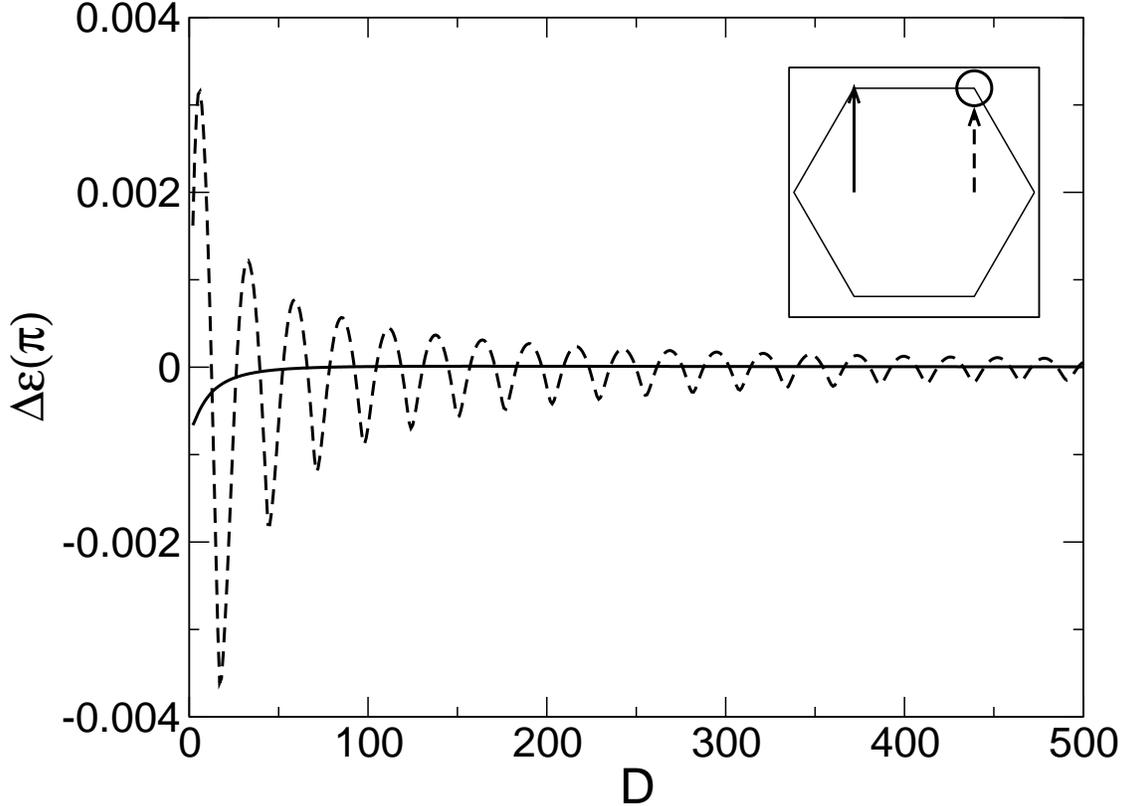}
\caption{Exchange coupling between Co adatoms as a function of the adatom separation
on a zigzag (6,0) nanotube (solid line). The dashed line is the coupling when the nanotube is 
slightly doped, corresponding here to a  small shift in its Fermi energy. The inset shows 
schematicaly the Fermi wave vector for pure (solid arrow) and doped (dashed arrow) tubes.
All coupling values are given in units of the nearest-neighbour 
electronic hopping $\gamma$.}
\label{zig1}
\end{figure}

We start by considering a metallic zigzag nanotube defined by the
indices (6,0). The solid line of figure \ref{zig1} shows the coupling
$\Delta {\cal E}(\pi)$ as a function of the distance $D$ between
adatoms along the axial direction of the tube. The parameters were
chosen to reproduce the adatom occupation of Co atoms. The first
noticeable feature in this result is the monotonicity of the
coupling. As a general physical principle, it is well know that the
indirect coupling between magnetic impurities embedded in a metallic
host oscillates as the impurities are moved further
apart.\cite{iec-review} At first glance, one may think that this
principle is not applicable to carbon nanotubes. However, as we show
below, this is not the case, the lack of oscillations being the result
of a commensurability effect.

On a three-dimensional system, in an analogy to the de Haas van Alphen
effect in metals, the oscillatory behaviour of the coupling is
determined by the topology of the bulk Fermi surface (FS) of the host
material. \cite{prl-bechara} More precisely, each FS extremum along a
specific direction contributes with one oscillatory component of a
definite period.  This period is determined by the caliper associated
with that extremum. For a CN, the Fermi ``surface'' is actually a
discrete set of points. Inspecting the zigzag's Fermi points one sees
that there is only one extremum, thus the coupling should oscillate
with one single period.  Nevertheless, the solid line in figure
\ref{zig1} is always negative indicating a preferential
antiferromagnetic alignment between the adatom moments. For infinitely
large separations, the coupling should vanish but it does so without
sign changes or any oscillations. This is because the period coming
from the caliper associated with that single extremum is actually
commensurate with the nearest-neighbour distance in the axial
direction. In other words, the expected oscillations are hidden by the
fact that the coupling is being probed at intervals that are of the
same lenght as the unit cell separation. To prove this point, we
introduce a slight shift in the Fermi level that would correspond to
either doping the nanotube or to the action of a gate voltage
affecting the occupation of the system. In this case, the Fermi
surface of the nanotube no longer consists of isolated points at the
corners of the Brillouin zone. The points in reciprocal space are now
located at the intersection between small circular-like curves centred
at the hexagonal corners of the Brillouin zone and the discrete
quantization lines along the $k_y$ direction, as shown in the inset of
figure \ref{zig1}. The inset displays the wave vectors that contribute
to the oscillatory coupling, which clearly shows that they no longer
coincide with the dimensions of the Brillouin zone. A small shift in
the Fermi level ($\Delta E_F = 0.06 \gamma$ in this case) is
sufficient to destroy the commensurability effect in the coupling, as
shown by the long-period oscillatory curve (dashed line) of figure
\ref{zig1}.

\begin{figure}
\includegraphics[width=0.9\textwidth,clip]{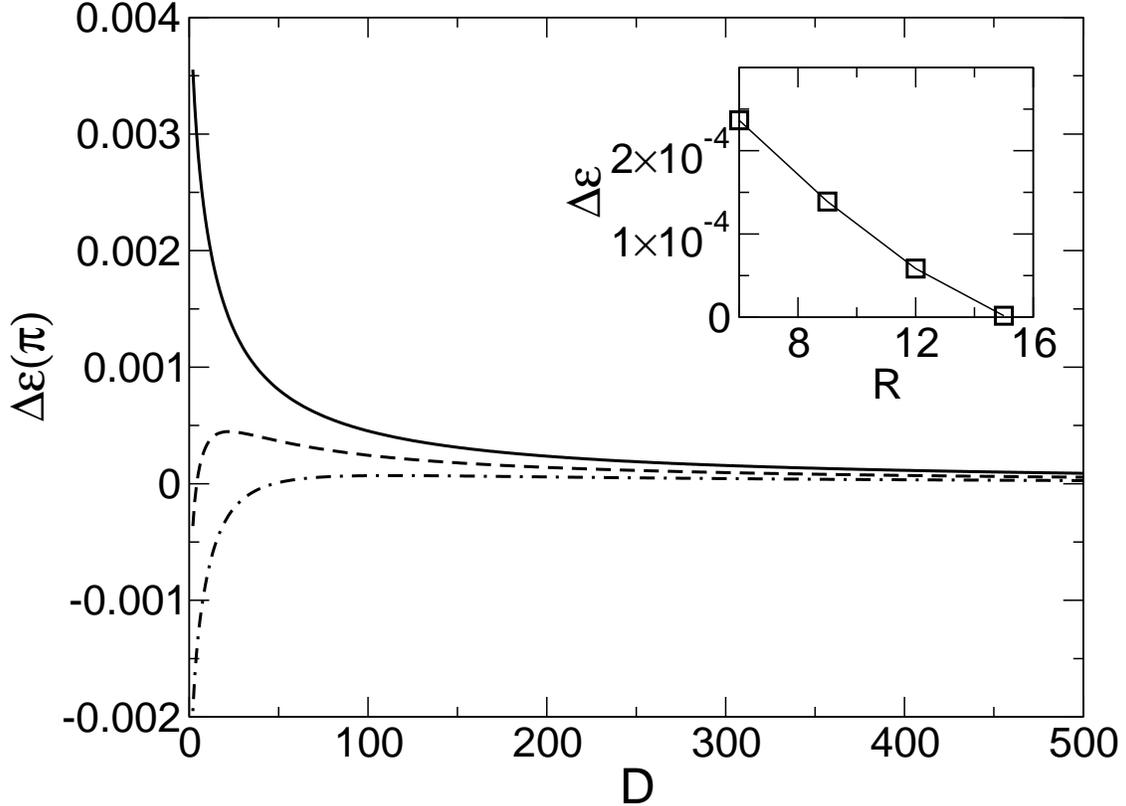}
\caption{Exchange coupling as a function of longitudinal separation $D$ between Ni adatoms
on (6,0) (solid line), (9,0) (dashed line) and (12,0) (dot-dashed line) nanotubes. The inset shows
how the coupling amplitude decays as a function of tube diameter. All coupling values are given 
in units of the nearest-neighbour electronic hopping $\gamma$.}
\label{zig2}
\end{figure}

The lack of oscillations in the indirect coupling is sufficiently
robust to appear with other magnetic adatoms. In fact, by changing the
parameters such that the d-band occupation of the adatoms corresponds
to that of Ni atoms, we find a similar behaviour for the coupling, the
only difference being the overall sign. This is illustrated by figure
\ref{zig2} in which the full line depicts the calculated coupling
between two Ni adatoms as a function of their longitudinal
separation. Rather than an antiferromagnetic alignment between the
moments, in this case we find that they tend to align in a
ferromagnetic fashion. Likewise, the monotonicity of the coupling
remains unnaffected by varying the tube diameter. This is confirmed by
the dashed and dot-dashed lines in figure \ref{zig2} in which the
coupling between Ni adatoms is plotted as a function of their
separation for a (9,0) and a (12,0) tube, respectively. The inset also
shows that the coupling decays with the tube diameter as $1/R$, where
$R$ is the nanotube radius. This is hardly surprising, bearing in mind
that the density of electronic states of a nanotube has a similar
behaviour and that the coupling is the result of an indirect coupling
mediated by the nanotube electrons.

\subsection{Zigzag (semiconductor)}

\begin{figure}
\includegraphics[width=0.9\textwidth]{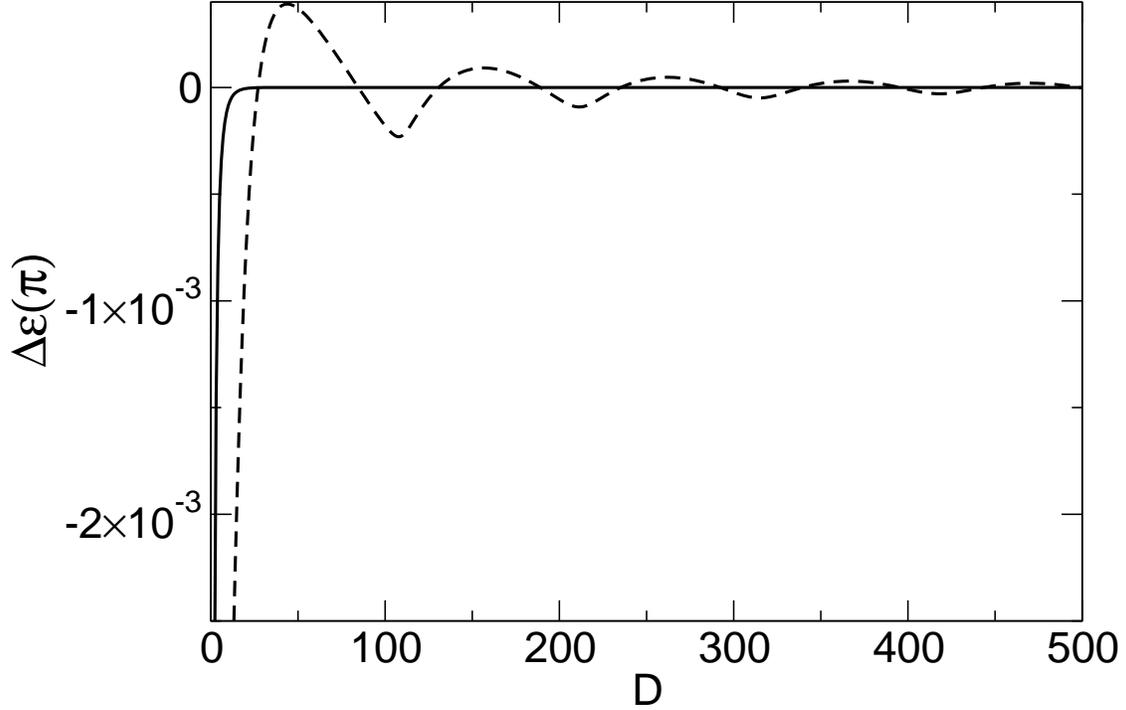}
\caption{Exponentially decaying exchange coupling between Co adatoms on a 
semiconducting (31,0) nanotube (solid line). A small amount of doping is enough 
to populate the tube's conduction band and give origin to an oscillating, long-ranged
exchange coupling (dashed line). All coupling values are given in units of the nearest-neighbour 
electronic hopping $\gamma$.}
\label{zig4}
\end{figure}

For the case of semiconducting tubes, the lack of extended states at
the Fermi level supresses the long-range character of the indirect
coupling, even in 1-dimension. In this case the coupling decays
exponentially with the adatom separation, as shown by the full line in
figure \ref{zig4} for a (31,0) nanotube. The choice of such a large
diameter tube is justified to illustrate the fact that a minute change
in the occupation is capable of moving the Fermi level into the zone
of extended states and producing a remarkable effect in the
coupling. The dashed line shows the coupling for a slightly doped
nanotube ($\Delta E_F = 0.06 \gamma$) displaying similar behaviour to
the metallic case, where a long-period long-ranged oscillatory
coupling is observed. From a virtually uncoupled configuration, the
coupling between adatom moments can be switched on by changing the
occupation of the nanotube. Consequently, by using gate voltages to
control the Fermi level of the system, in principle one can reversibly
turn magnetically uncoupled adatoms into fully aligned magnetic
moments, a promising feature for use in spintronic devices.

\subsection{Armchair}

\begin{figure}
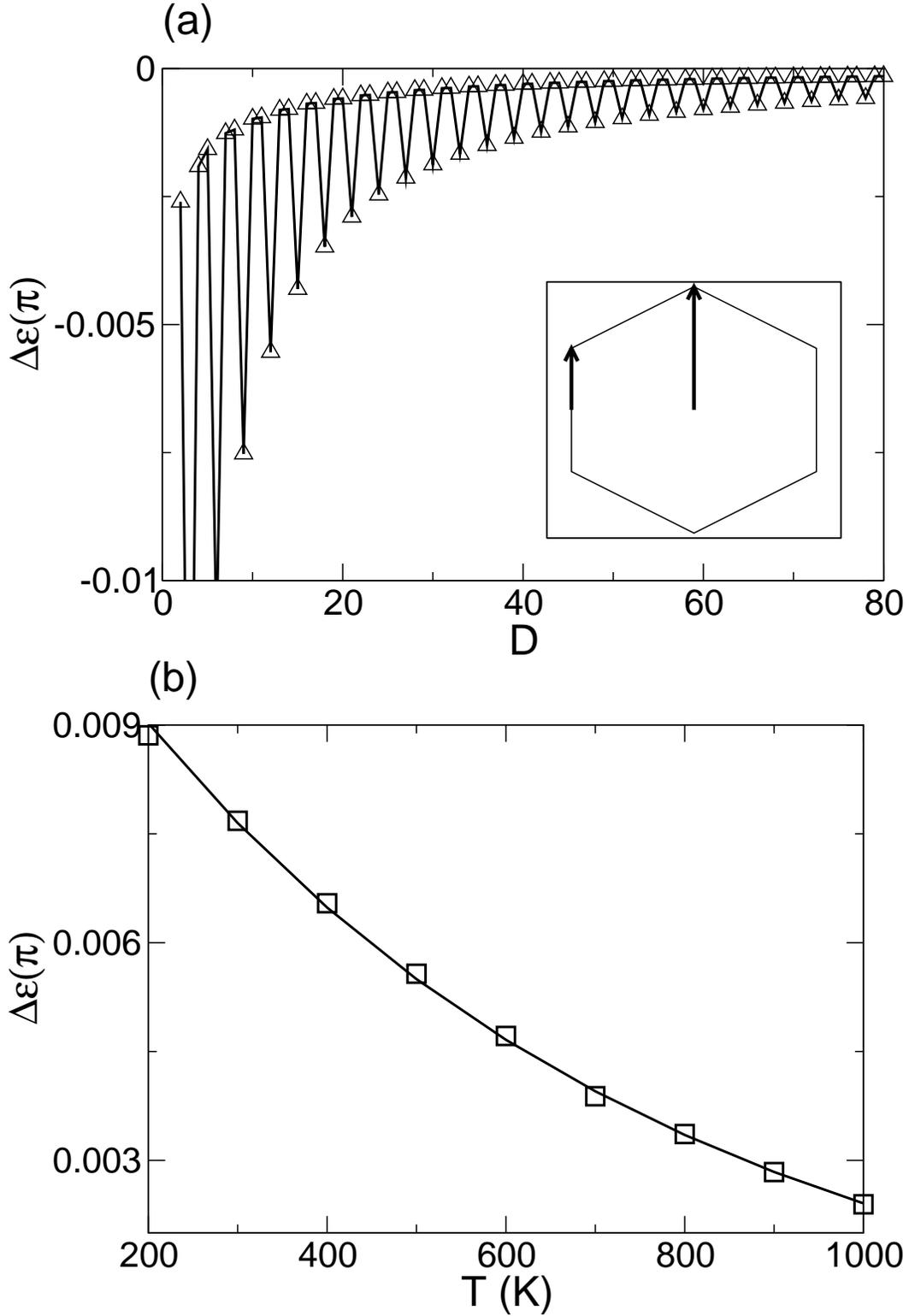

\centerline{\includegraphics[width=0.85\textwidth,clip]{fig5a.eps}}
\centerline{\includegraphics[width=0.85\textwidth,clip]{fig5b.eps}}
\caption{a) Exchange coupling as a function of longitudinal separation $D$ between two 
Co adatoms on a (6,6) armchair nanotube. The inset shows the wave vectors contributing
to the coupling with different oscillation periods. b) Amplitude of the exchange coupling 
as a function of temperature. All coupling values are given in units of the nearest-neighbour 
electronic hopping $\gamma$.}
\label{arm1}
\end{figure}

For the armchair tubes, there are two contributory wave vectors
extracted from their Fermi surface, and as such there should be two
oscillatory components in the exchange coupling. These are identified
in the inset of figure \ref{arm1} together with the coupling as a
function of the adatom separation in the main graph. The full line
corresponds to the coupling evaluated at $T = 0$ and the dashed line
shows the coupling at room temperature. Once again, one of the periods
is commensurate with the interatomic distance, which makes the
coupling to oscillate with a non-commensurate period superimposed to a
steady decay. The coupling at finite temperature tends to decay
faster, which can be explained by the fact that the thermal energy
partially suppresses the propagators between adatoms, thus reducing
the efficiency with which the electronic carriers transport the
magnetic information back and forth.

\begin{figure}
\includegraphics[width=0.9\textwidth,clip]{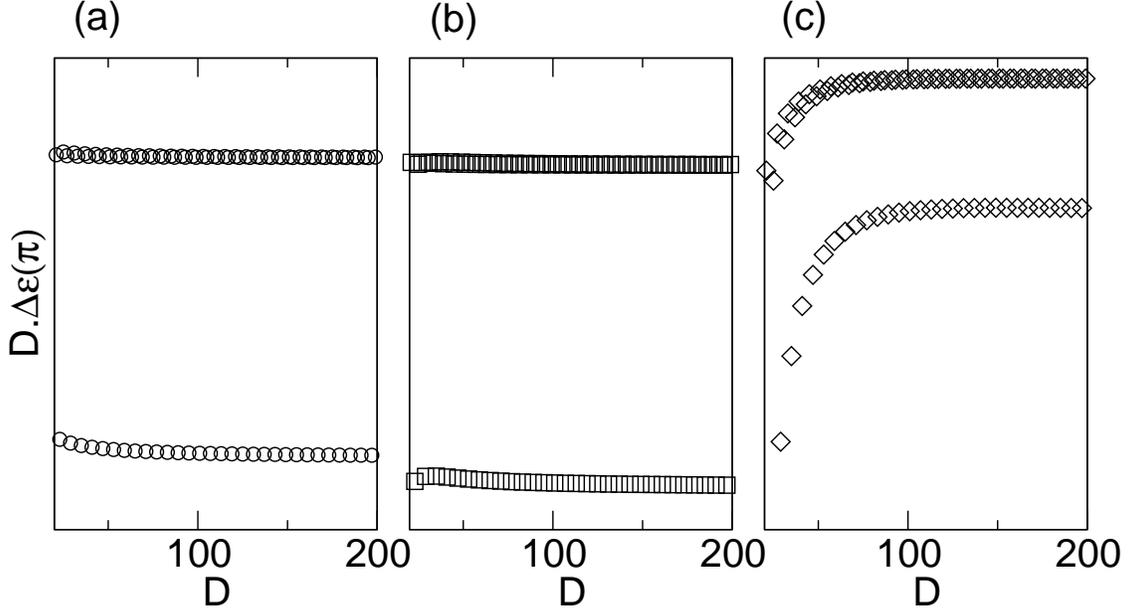}
\caption{Exchange coupling $\Delta{\cal E}(\pi)$ multiplied by the distance between impurities $D$,
as a function of $D$, for a) (6,6), b) (24,24)  and c) (48,48) armchair nanotubes. It is evident that in the
(6,6) and (24,24) nanotubes the coupling decays as $1/D$, whereas in the (48,48) tube it decays faster.
All coupling values are given in units of the nearest-neighbour 
electronic hopping $\gamma$.}
\label{imp_decay}
\end{figure}

Concerning the rate of decay of the coupling as a function of the
adatom separation, all examples above decay slowly in the asymptotic limit ($D>>1$). 
It is known that in one-dimensional structures the coupling decays approximately
as $1/D$ for $D>>1$. This is what we observe approximately when impurities embedded 
in the nanotube are considered, instead of adatoms. As the tube diameter 
increases, the system approaches a 2-dimensional-like structure and the coupling 
tends to decay a little faster. This is displayed in Figure \ref{imp_decay} showing the 
approximate decaying rate as a function of the tube diameter for armchair
structures. The coupling between adatoms has a more complicated asymptotic 
behaviour. For most of the range of $D$ values we investigated it was not possible
to identify a unique decay rate. In the regions where the decay rate is well-defined
it may be even slower than $1/D$. Differences between the coupling decay rates for 
adatoms and embedded impurities have been reported in other systems \cite{sonia}.

It is worth mentioning that such a slow decaying rate in
the coupling leads to a long-ranged magnetic interaction that, when
combined with the lack of oscillations introduced by the
commensurability effect, can be potentially used to induce ordered
magnetic structures on the surface of a carbon nanotube.

\section{Conclusions}

In summary, we have presented expressions for the indirect exchange
coupling between magnetic adatoms mediated by the conduction electrons
of carbon nanotubes. It is obvious from the expressions that the
coupling, written in terms of a few matrix elements of the
single-particle Green functions, depends on the electronic carriers
transporting the magnetic information back and forth to both
adatoms. We show that, contrary to expectations, we find a monotonic
behaviour in the coupling between magnetic adatoms adhered to the
surface of metallic zigzag nanotubes. Rather than an intrinsic
property, we have shown that the lack of usual oscillations results
from a commensurability effect. In other words, the oscillations are
hidden by the fact that the contributory wave vectors have the same
size as the Brillouin zone, forcing the periods to coincide with the
atomic spacing. In this commensurate regime the coupling does not
change sign but decays monotonically. The overall sign of this
monotonic coupling depends on the nature of the adatoms attached to
the nanotube. In the case of positive coupling, a FM alignment between
the moments persists as we move the adatoms apart, where the coupling
magnitude decays rather slowly as approximately $1/D$. Such a long
range interaction may be responsible for inducing magnetic order in
the nanotube.

The coupling across semiconducting tubes is exponentially damped,
which results from the lack of extended states to transport the
magnetic information between the adatoms. By moving the occupation
into the conduction band we show a dramatic change in the coupling,
departing from a configuration with random alignment between magnetic
moments into another that is highly ordered due to the predominantly
positive long-ranged long-period oscillatory coupling. We argue that a
controlled gate voltage can potentially tune the ocuppation in such a
way that the magnetic interaction can be switched on and off in a
reversible fashion.

Oscillatory coupling is found on armchair metallic tubes. In this
case, two periods are identified, one commensurate and another
incommensurate with the lattice. The resulting coupling consists of a
monotonically decaying contribution superimposed to a $1/D$-modulated
oscillatory coupling. Concerning the dependence of the coupling on
other parameters, its diameter dependence follows the $1/R$-behaviour
displayed by the density of states. As expected, the coupling decays
with temperature but not as fast as the coupling between moments
embedded in 3-dimensional structures.

Finally, it is worth emphasizing that by controlling the alignment of
magnetic structures attached to the surface of carbon nanotubes one
can potentially select the way in which the systems will respond to
magnetic excitations. Moreover, bearing in mind that the transport
properties of magnetically doped structures are sensitive to the
alignment of their magnetizations, the understanding of how separate
moments are coupled across nanotubes could be a major step forward in
the search for and control of spin-valve effects in those structures.

\section{Acknowledgments}
Useful discussions with R. B. Muniz are gratefully acknowledged. The
authors thank Science Foundation Ireland for financial
support. A. T. Costa-Jr is grateful to the Brazilian agencies CNPq,
FAPEMIG and FINEP for financial support.


\begin{thebibliography}{99}
\bibitem{comment} The difference to the original expression for the
indirect coupling derived in reference \onlinecite{castro94} is the
absence of in-plane translational invariance that exists in the case
of multilayered structures.
\end{thebibliography}
\end{document}